\documentclass[aps,prb,twocolumn,groupedaddress,amsmath,amssymb]{revtex4}

\usepackage{graphicx}
\usepackage{dcolumn}
\usepackage{bm}

\newcommand{\pdag}{{\phantom{\dagger}}}

\newcommand{\CeAu}{CeCu$_{6-x}$Au$_x$}

\newcommand{\YbRhSi}{YbRh$_2$Si$_2$}

\begin{document}

\title{
Kondo volume collapse, Kondo breakdown, and Fermi surface transitions\\
in heavy-fermion metals
}

\author{Andreas Hackl}
\email[]{ah@thp.uni-koeln.de}
\author{Matthias Vojta}

\affiliation{
Institut f\"ur Theoretische Physik, Universit\"at zu K\"oln,
Z\"ulpicher Stra\ss e 77, 50937 K\"oln, Germany
}

\date{\today}

\begin{abstract}
The unconventional critical behavior near magnetic quantum phase transitions
in various heavy-fermion metals, apparently inconsistent with the standard
spin-density-wave scenario, has triggered proposals on the breakdown
of the Kondo effect at the critical point.
Here we investigate, within one specific scenario,
the fate of such a zero-temperature transition
upon coupling of the electronic to lattice degrees of freedom.
We study a Kondo-Heisenberg model with volume-dependent
Kondo coupling -- this model displays both Kondo volume collapse and
Kondo-breakdown transitions, as well as Lifshitz transitions
associated with a change of the Fermi-surface topology.
Within a large-$N$ treatment, we find that the Lifshitz transition
tends to merge with the Kondo volume collapse and hence becomes first order,
whereas the Kondo breakdown transition remains of second order except for
very soft lattices.
Interesting physics emerges at the zero-temperature endpoint of
the Kondo volume collapse:
For electrons in two space dimensions, this endpoint is located {\em at} the Lifshitz
line for a large range of parameters, thus two continuous quantum phase transitions
coincide without fine tuning.
We analyze the effective Landau theory for such a situation
and discuss critical exponents.
Finally, we relate our findings to current heavy-fermion experiments.
\end{abstract}

\pacs{75.20.Hr, 75.30.Mb, 71.10.Hf}

\maketitle


\section{Introduction}
\label{intro}

The past decade has seen growing interest in continuous quantum phase transitions.
The associated finite-temperature quantum critical regime
is characterized by unconventional dynamics and thermodynamics;
in particular, quantum criticality can profoundly modify the metallic properties of
both $d$- and $f$-electron materials.
Heavy-electron metals have played a
particularly important role in the study of quantum criticality, for these materials lie
at the brink of antiferromagnetic (AF) instabilities, and they are readily tuned to an AF
quantum critical point (QCP). The occurrence of this QCP is commonly attributed  to a
competition between the Kondo screening of the local moments and a tendency to AF
ordering due to an inter-moment exchange interaction,
as first proposed by Doniach.\cite{doniach}
Interestingly, various properties near the critical points of materials like
\CeAu\ and \YbRhSi\ appear to be inconsistent with the
standard spin-density-wave (or Landau-Ginzburg-Wilson) scenario of the magnetic
phase transition.\cite{hvl}
Most prominent here are the singular behavior of the specific-heat coefficient
and $E/T$-scaling in inelastic neutron scattering experiments.\cite{schroeder}
These have led to suggestions that the critical point is not simply a magnetic
instability of fermionic quasiparticles, but that in addition the Kondo effect,
being responsible for the formation of these quasiparticles, breaks
down.\cite{si,coleman,senthil,senthil2}

Assuming that the Kondo breakdown is the primary source of the critical singularities,
Senthil {\em et al.}\cite{senthil,senthil2} have suggested a model where the Kondo breakdown
occurs as a topological quantum phase transition between two paramagnetic phases,
namely a heavy Fermi liquid (FL) and a so-called fractionalized Fermi liquid (FL$^\ast$),
and ordered magnetism is regarded as an epiphenomenon.
As in the Doniach picture, the formal control parameter of the transition is the ratio $T_K/J_H$
between Kondo temperature and Heisenberg exchange of local moments.
It is, however, assumed that the local moments form a spin-liquid state without
broken symmetries (instead of the standard local-moment antiferromagnetism)
in the regime $T_K \ll J_H$.
Such a spin liquid may result from strong quantum fluctuation, induced
e.g. by frustrated RKKY or exchange interactions between the local moments.\cite{sl}

As phase transitions in heavy-fermion metals are often tuned by external or chemical pressure,
the coupling of electronic to lattice degrees of freedom can play a non-trivial role
for the overall shape of the phase diagram. This aspect motivates our paper.
A well known but spectacular example are the pressure-induced volume-collapse transitions
in the trivalent rare earth metals,\cite{mahan} like the transition between $\alpha$ and
$\gamma$ Ce.
In this example, a line of first-order transitions is found in the
pressure--temperature plane where the unit cell reduces its volume by about $15 \%$.
An order of magnitude change of the Kondo temperature accompanies this transition.
These aspects have led to the Kondo volume-collapse model,\cite{allen,cyrot}
which explains the transition based on a volume dependence of the Kondo exchange
coupling.

In this paper, we shall study a model which contains the ingredients both
for the Kondo-breakdown quantum phase transition as well as
for Kondo volume-collapse physics,
with the goal of obtaining a unified phase diagram of these phenomena.
As both transitions can in principle be tuned by pressure, with the
inter-atomic distances strongly influencing the Kondo energy scales,
an intricate interplay of electronic and lattice degrees of freedom may
be expected.
Of particular interest will be whether the Kondo-breakdown transition of
the type proposed in Ref.~\onlinecite{senthil2} survives
as a second-order quantum transition once a coupling to the lattice
is taken into account.

\subsection{Summary of results}

Our microscopic analysis is based on a fermionic large-$N$ treatment of a Kondo-Heisenberg
model in both two and three spatial dimensions,
with the Kondo exchange being linearly coupled to static lattice distortions.
In the scenario of Ref.~\onlinecite{senthil2}, the Kondo-breakdown transition
between the FL and U(1) FL$^\ast$ phases occurs in the presence of two Fermi surface sheets;
as a result, a Lifshitz transition generically occurs upon tuning the ratio $T_K/J_H$
within the heavy Fermi-liquid phase. At such a Lifshitz transition,\cite{lifshitz}
well known from weakly interacting Fermi systems, a topological splitting of the
Fermi surface occurs, here leading to a transition between Fermi liquids
with one and two Fermi sheets (dubbed FL$_1$ and FL$_2$, respectively).

The physical behavior of our system depends on both the ratio $T_K/J_H$
and the electron--lattice coupling (or the lattice stiffness).
Zero-temperature phase diagrams, as function of pressure (which linearly couples to $T_K$)
and $J_H$, are shown below in Figs.~\ref{2dphase}, \ref{3dphase}.
Interestingly, the above-mentioned Lifshitz transition is easily driven first order upon
inclusion of lattice degrees of freedom, and is then accompanied by a jump in lattice
volume and Kondo temperature.
In contrast, the Kondo breakdown transition FL$_2$--FL$^\ast$ remains of second order
except for very soft lattices -- in the latter case, the Fermi-liquid breakdown
occurs from FL$_1$ to FL$^\ast$ as a strong direct first-order transition.
For small $J_H$, the behavior is dominated by the interplay of Kondo and lattice
physics, and soft lattices admit a separate Kondo volume collapse
transition (as in earlier work).
This first-order transition merges with the Lifshitz transition at larger $J_H$.

For a large range of lattice stiffness values, a zero-temperature endpoint of the
Kondo volume collapse occurs in the $p$--$J_H$ phase diagram.
Remarkably, in two spatial dimensions this endpoint is quite generically located
{\it on} the Lifshitz transition line without further fine tuning.
This coincidence of two continuous transitions can be further analyzed using
an effective theory for coupled Lifshitz and lattice order parameters.
It turns out that the electron--lattice coupling leads to non-perturbative
effects which are ultimately responsible for the ``non-Landau'' transition
behavior.

\subsection{Relation to earlier work}

Our work merges ideas on the Kondo volume collapse and those of
quantum critical behavior near magnetic phase transitions in heavy-fermion compounds.
Furthermore, earlier works on quantum Lifshitz transitions play a role as well.

Historically, two main theoretical approaches have been advanced to describe the
$\gamma \rightarrow \alpha$ transition of elemental Cerium, namely the
Mott transition scenario\cite{johansson} and the Kondo volume collapse
model.\cite{allen, cyrot}
Common to both scenarios is that the $f$ electrons are more localized in
the $\gamma$ phase.
More recently, the modern dynamical mean-field theory (DMFT) in combination with
realistic band structure calculations (LDA+DMFT) was employed, and reasonable agreement
with experimental data has been achieved.\cite{held}
The relation between the Kondo volume collapse and quantum criticality was only
discussed recently in Ref. \onlinecite{schmalian},
where the situation of the endpoint suppressed to $T=0$ was considered.
This work did not include competing exchange interactions between local moments, and
hence did not contain the physics of magnetic or Kondo criticality.

The breakdown of the lattice Kondo effect at the QCP in heavy fermions was proposed
in Refs.~\onlinecite{coleman,si}. In particular, the approach of Si {\em et al.}\cite{si}
used an extension of DMFT to obtain a scenario of ``local quantum criticality'',
where the onset of magnetism in two dimensions suppresses Kondo screening right at
criticality.
A somewhat different viewpoint was put forward in Refs.~\onlinecite{senthil,senthil2},
where Kondo criticality was discussed in the absence of ordered magnetism.
This required the existence of a novel phase, a fractionalized Fermi liquid (FL$^\ast$),
which is a metallic spin-liquid phase with emergent gauge structure.
Central to the Kondo-breakdown transition\cite{senthil,senthil2} is that the $f$
electron subsystem becomes insulating, hence the transition can be described
as an orbital-selective Mott transition.\cite{osmott1,paul}
Ref.~\onlinecite{senthil2} studied the mean-field theory to be used below,
but the coupling to lattice degrees of freedom was not discussed,
and the Lifshitz transition in the FL phase (which will turn out to be important in our
analysis) was mentioned, but not investigated in detail.
Similar mean-field theories also appear in Refs.~\onlinecite{burdin,coqblin},
but not with focus on the Kondo breakdown transition.

Lifshitz transitions in correlated electron systems, with topological changes of the
Fermi surface, have been discussed at length in Ref. \onlinecite{imada}.
However, this analysis did not contain the coupling of the electrons to other
collective degrees of freedom.

\subsection{Outline}

The body of this paper is organized as follows:
In Sec.~\ref{sec:model} we formulate the Kondo-Heisenberg model supplemented by
electron--lattice coupling and discuss its large-$N$ solution and the
resulting zero-temperature phases.
Fully self-consistent numerical calculations of the zero-temperature phase diagrams in
the parameter regime of pressure and inter-moment exchange are presented afterwards.
Particular attention is paid on the role of dimensionality and geometry of the underlying
lattice.
We discuss the existence of discontinuous volume collapse transitions that occur in
a wide range of our numerically obtained phase diagrams.
In Sec.~\ref{sec:landau} we focus on the interplay of Lifshitz and volume-collapse
transitions in the framework on an effective Landau theory.
This analysis is in agreement with the mean-field results for the Kondo lattice,
and furthermore shows that, for two-dimensional (2d) electrons,
the quantum critical endpoint of the first-order
volume collapse transition tends to coincide with the second-order Lifshitz
transition.
Fluctuations beyond mean-field theory are discussed in Sec. \ref{sec:fluct},
and a brief summary and outlook concludes the paper.


\section{Microscopic Kondo-lattice theory}
\label{sec:model}

The starting point of our analysis is the standard Anderson lattice
model, describing conduction electrons on a lattice which hybridize with local
atomic $f$-orbitals on the lattice sites.
\begin{eqnarray}
{\cal H} &=&
\sum_{k\sigma} \epsilon_k c_{k\sigma}^\dagger c_{k\sigma}^\pdag + \sum_{i\sigma} \epsilon_f^{0} f_{i\sigma}^\dagger
f_{i\sigma}^\pdag + U \sum_i n_{i\uparrow }^f n_{i\downarrow }^f \nonumber\\ &+& \frac{1}{\sqrt{\mathcal{N}}}\sum_{k i\sigma} (V_k e^{-ikR_i} c_{k\sigma}^\dagger f_{i\sigma}^\pdag + H.c.)
\label{pam}
\end{eqnarray}
The first term describes conduction electrons with band filling $n_c$,
moving on a two-dimensional (2d) or three-dimensional (3d) regular lattice
(the former case applies to layered systems with weak electronic inter-layer coupling),
$\varepsilon_f^0$ ($U$) is the bare $f$ electron energy (Coulomb repulsion),
$V_k$ the hybridization matrix element,
and $\mathcal{N}$ the number of unit cells.

Let us now discuss the influence of volume changes induced by external pressure.\cite{luethi}
Experimental data on Kondo volume collapse transitions in Ce shows that,
throughout the transition,
the $f$ level occupation and the $f$ electron levels show only modest
changes.\cite{foot1}
In contrast, the Kondo temperature changes rapidly with volume by orders of magnitude.
These results motivate us to neglect any volume dependence of the Anderson lattice model
other than that of the hybridization matrix elements.
(Note that the Kondo temperature depends exponentially on both the
hybridization and the conduction-electron bandwidth.
Experimentally, both will be influenced by volume changes.
For our qualitative considerations, taking into account the volume
dependence of one is sufficient.)

For most of the paper, we will treat the lattice degrees of freedom classically,
parameterized by the strain tensor $\hat{\epsilon}$.
At a transition between two iso-structural states a suitable order parameter is the
trace $\epsilon$ of the strain tensor,
\begin{equation}
\epsilon \equiv {\rm Tr}\,\hat{\epsilon} =\frac{V-V_0}{V_0} ,
\label{defeps}
\end{equation}
where $V_0$ is a reference volume.
In analogy to Ref. \onlinecite{schmalian}, the hybridization matrix elements are
allowed to depend on the local strain $\epsilon_i$ by a linearized change of the overlap
integral $V_k= \langle k|\hat{V}|f\rangle$ between a Bloch electron and a local
$f$ electron with its atomic potential $\hat{V}$.
Assuming a local hybridization, $V_k\equiv V$,
we account for the lattice coupling be replacing
${\cal H} \rightarrow {\cal H} +{\cal H}_c$ with
\begin{equation}
{\cal H}_c=\gamma V \sum_{i\sigma} \epsilon_i (f_{i\sigma}^\dagger c_{i\sigma}^\pdag +H.c) + \frac{B}{2}
\frac{V_0}{\mathcal{N}}\sum_{i} \epsilon_i^2
\label{strain}
\end{equation}
where $\gamma$ is the coefficient of the assumed linear local strain dependence of the
hybridization.
The last term captures the deformation energy, with $B$ being the bulk modulus.
For instance, in a cubic system, $B=\frac{1}{3}(c_{11}^0+ 2 c_{12}^0)$,
with bare elastic constants $c_{11}^0$ and $c_{12}^0$.\cite{luethi}

Because of the exponential variation of $T_K$ with $J_K$, moderate variations of the
Kondo coupling $J_K$ with volume cause a strong volume dependence of the Kondo
temperature.
This, in turn, yields a nonlinear equation of state $p(V)=-\frac{\partial
F}{\partial V}|_T=-\frac{1}{V_0}\frac{\partial F}{\partial \epsilon}|_T$.
It has been shown\cite{allen} that models of this type yield nonlinear $p$-$V$ isotherms
similar to the van der Waals theory of the liquid--gas transition.
The phase diagram of the model from Eq. (\ref{pam}) and (\ref{strain}) has been discussed
in Ref.~\onlinecite{schmalian}, and a zero-temperature volume collapse within
the heavy-Fermi liquid phase has been found to occur below a critical value $B^\star$ of
the bulk modulus.

In this paper, we shall discuss the interplay of volume collapse transitions
with a possible breakdown of Kondo screening due to competing inter-moment exchange.
To simplify the following approximate treatment, we shall
supplement our model by an explicit Heisenberg-type exchange interaction
between the local moments,\cite{senthil,senthil2}
and we shall pass from the Anderson to the Kondo model.\cite{schrieffer}
The latter step, justified in the Kondo limit
$\frac{V^2}{\epsilon_f + U},\frac{V^2}{\epsilon_f}\rightarrow 0$,\cite{foot2}
neglects possible charge fluctuations which we assume to be of minor
influence for the structure of phase diagram.
Note that we only fix $f$ valence (which is known to not vary drastically across
a volume collapse transition) to unity, whereas we still allow for a transition
from ``itinerant'' to ``localized'' $f$ electrons.\cite{senthil2}
We arrive at a Kondo-Heisenberg model of the form
\begin{eqnarray}
{\cal H}_{\rm KHM} &=&
\sum_k \epsilon_k c_{k \sigma}^\dagger c_{k\sigma}^\pdag +
 J_K(1+\gamma \epsilon)^2 \sum_r \vec{S}_r \cdot \vec{s}_r \nonumber\\
&&- \frac{N}{4}J_K(1+\gamma \epsilon)^2 +
\frac{B}{2} V_0 \epsilon^2 \nonumber\\
&&+ J_H \sum_{<rr^\prime>} \vec{S}_r \cdot \vec{S}_{r^\prime}
\label{khm}
\end{eqnarray}
where $\vec{s}_r=\frac{1}{2}c_{r\sigma}^\dagger \vec{\sigma}_{\sigma \sigma^\prime} c_{r
\sigma^\prime}^\pdag$ is the local spin density of the conduction electrons.
We neglect the influence of lattice distortions on inter-orbital magnetic exchange,\cite{luethi}
for two reasons:
(i) The exponential dependence of Kondo temperature on the strain likely dominates
over other instances of electron--lattice couplings.
(ii) The physics is primarily influenced by the ratio $T_K/J_H$, and for qualitative
considerations the strain dependence of one of the parameters is sufficient.

In the following, the zero-temperature phase diagram of the model ${\cal H}_{\rm KHM}$ \eqref{khm}
will be determined within a fermionic mean-field approach,\cite{senthil2}
which allows to capture both the volume collapse and the Kondo breakdown transitions.
An implicit assumption is that quantum fluctuations of the local moments
are large, such that a spin liquid replaces ordered magnetism for $T_K \ll J_H$.
Technically, this assumption determines the choice of the mean-field decoupling of $J_H$
to be described below.

\subsection{Fermionic large-$N$ theory}

The model \eqref{khm} cannot be solved analytically as it stands.
Instead, much progress can be made by solving it in the large-$N$ approach which has been
extensively used in the study of the Kondo lattice and also allows to deal with the
magnetic exchange term in Eq. \eqref{khm}.\cite{burdin,coqblin}

Using a fermionic representation of the local moments allows us to capture phases
both with and without Kondo screening.
Importantly, for dominant exchange interaction $J_H$, one obtains
a spin-liquid or valence-bond ground-state,\cite{senthil,senthil2}
which replaces the metallic antiferromagnet conventionally found
in this regime.\cite{doniach}
Here, we will restrict our attention to phases without broken lattice
symmetries, hence tuning the ratio $J_H/T_K$ drives a transition between
a heavy Fermi liquid (FL) and a fractionalized Fermi liquid (FL$^\ast$) --
in the latter phase the conduction-electron and local-moment subsystems are
decoupled, and the $f$ electrons form a Mott-insulating spin liquid state.
The quantum phase transition between FL and FL$^\ast$ is accompanied
by a jump in the Fermi volume.\cite{senthil2}

The large-$N$ approach to the Kondo-Heisenberg model, Eq. \eqref{khm}, extends the
local-moment symmetry to the SU($N$) symmetry group.
The $N^2-1$ components $(S_{\sigma \sigma^\prime}^r)_{1\leq \sigma,\sigma^\prime\leq N}$
of a local moment $\vec{S}(r)$ are
represented in terms of neutral fermions $f_{r\sigma}$ with a local constraint:
\begin{equation}
S_{\sigma \sigma^\prime}^r= f_{r\sigma}^\dagger f_{r \sigma^\prime}^\pdag-\frac{\delta_{\sigma \sigma^\prime}}{2}, ~~~ \sum_{\sigma=1}^N
f_{r\sigma}^\dagger f_{r\sigma}^\pdag = \frac{N}{2}
\label{pseudof}
\end{equation}

The interaction terms of Eq. \eqref{khm} now become quartic. These are decoupled
introducing Hubbard-Stratonovich fields $b_r(\tau)$ conjugate to $\sum_\sigma
f_{r\sigma}^\dagger c_{r\sigma}^\pdag$ and $\chi_{ij}(\tau)$ conjugate to $\sum_\sigma
f_{i\sigma }^\dagger f_{j\sigma }^\pdag$ and the constraint is enforced via
Lagrange multipliers $\lambda_r(\tau)$. The action corresponding to the
representation (\ref{pseudof}) of Eq. \eqref{khm} becomes extensive in $N$ by rescaling
the coupling constants as $J_K\rightarrow \frac{J_K}{N}$, $J_H\rightarrow \frac{J_H}{N}$.
In the $N\rightarrow \infty$ limit the physics is controlled by a saddle point at which
the Bose fields condense.
We assume a spatially homogeneous mean-field state\cite{foot3,foot3b} where $b_r$, $\lambda_r$,
and $\chi_{ij}$ take constant values $b_0$, $\lambda$, and $\chi_0$.
The mean-field Hamiltonian takes the form\cite{foot4}
\begin{eqnarray}
{\cal H}_{\text{mf}} &=& \sum_{k\sigma} (\varepsilon_k -\mu_c(\epsilon) )c_{k\sigma}^\dagger c_{k\sigma}^\pdag \nonumber\\
&&- \sum_{k\sigma}
\left(\lambda + 2 \chi_0 \sum_{\vec{r}=\text{n.n.}} e^{i\vec{r}\cdot\vec{k}} \right)
f_{k\sigma}^\dagger f_{k \sigma}^\pdag \nonumber\\
&&- b_0 (1+\gamma \epsilon)^2\sum_k (c_{k\sigma}^\dagger f_{k\sigma}^\pdag +h.c.) \nonumber\\
&&-\frac{J_K}{4}(1+\gamma \epsilon)^2 +V_0\frac{B}{2}\epsilon^2
\label{hmeanfield}
\end{eqnarray}
where we have dropped additional constants. The saddle point equations can be written in
the following compact form:
\begin{equation}
\left\{
\begin{array}{c}
\frac{b_0(1+\gamma\epsilon)^2}{J_K}\\
\frac{\chi_0}{J_H}\\
1\\
n_c\\
\end{array}
\right\}= \sum_k \left\{
\begin{array}{l}
G_{fc}(k,\tau=0^-)\\
G_f (k,\tau=0^-) (\varepsilon_{kf}-\lambda)\\
G_{f}(k,\tau=0^-)\\
G_c (k,\tau=0^-)\\
\end{array}
\right\}
\end{equation}
where $\varepsilon_{kf}=\lambda + 2 \chi_0 \sum_{\vec{r}=\text{n.n.}}
e^{i\vec{r}\cdot\vec{k}}$ is the dispersion of the $f$-fermions and $G_c$, $G_f$ and
$G_{fc}$ are the full conduction-electron, $f$-electron and mixed Green functions, respectively,
obtained from the Matsubara Green functions:
\begin{eqnarray}
&& G_c(i\omega_n,k) =  G_c^0(i\omega_n + \mu_c -b_0^2 G_f^0(i\omega_n ,k ),k) \nonumber\\
&& G_f(i\omega_n,k) =  G_f^0(i\omega_n  -b_0^2 G_c^0(i\omega_n ,k ),k) \nonumber\\
&& G_{fc}(i\omega_n,k) =  \frac{b_0 G_f^0(i\omega_n,k) G_c^0(i\omega_n,k ) }{1-b_0^2G_f^0(i\omega_n,k) G_c^0(i\omega_n,k )}
\label{largen}
\end{eqnarray}
Here $G_c^0(i\omega_n,k)=(i \omega_n -\varepsilon_k)^{-1}$ and $G_f^0(i\omega_n,k)=
(i\omega_n-\varepsilon_{kf})^{-1}$ are the non-interacting conduction and
$f$ electron Green functions.

The equilibrium lattice strain $\epsilon$ minimizes the free-enthalpy function
\begin{equation}
G(\epsilon)=F(\epsilon) + pV_0\epsilon
\end{equation}
what leads to the additional self-consistency condition
\begin{eqnarray}
\epsilon &=&\biggl( \frac{B V_0}{\gamma^3}-\frac{pV_0}{\gamma^2} \biggr)  \biggl( (\frac{3}{2}-n_c)J_K  \nonumber\\
&+&\frac{8b_0^2}{J_K}  + \frac{BV_0}{\gamma^2} \biggr)^{-1} - \gamma^{-1}
\label{epsilon}
\end{eqnarray}
which completes the set of mean-field equations together with Eq. (\ref{largen}).

\subsection{Zero-temperature phases}

In Eq. (\ref{hmeanfield}), the effect of the strain $\epsilon$ is just to renormalize
chemical potentials and the Kondo coupling $J_K$. Thus, there are two qualitatively
different zero-temperature phases of our mean-field model which where already derived in
Ref. \onlinecite{senthil2}.
On the one hand, there is the usual Fermi-liquid (FL) phase when $b_0$,
$\chi_0$, $\lambda$ are all nonzero. (Note that $b_0 \neq 0$ implies $\chi_0 \neq 0$.)
One the other hand, the Kondo hybridization may be zero, $b_0=0$,
but $\chi_0\neq 0$ (and $\lambda=0$) -- this is the FL$^\ast$ phase.
In this mean-field state the conduction electrons are decoupled from the local moments
and form a {\it small\/}\cite{senthil2} Fermi surface.
The local-moment system is described as a U(1) spin liquid with a Fermi surface of neutral
spinons. We expect that $\chi_0 \sim J_H$.
At finite temperatures the phase transition becomes a crossover -- this feature is not
captured by mean-field theory but requires to consider the coupling to the compact
U(1) gauge field.\cite{senthil2}
(Spin liquid states where the gauge group is broken down to Z$_2$ are also possible:
Those lead to superconductivity masking the FL--FL$^\ast$ transition\cite{senthil}
and shall not be considered here.)

We note that the large-$N$ mean-field approach can be expected to be qualitatively
correct at zero temperature: The FL phase is stable w.r.t. fluctuations, as is
FL$^\ast$, {\em provided that} the U(1) gauge field is in a deconfined phase.
For a further discussion of fluctuation effects we refer the reader to
Ref.~\onlinecite{senthil2}.

The mean-field Hamiltonian can be interpreted as a model of non-interacting
quasiparticles with dispersions
\begin{equation}
E_{k}^\pm=
\frac{\varepsilon_k+ \varepsilon_{kf}}{2} \pm \sqrt{ \frac{(\varepsilon_k - \varepsilon_{kf})^2}{2} + 4(1+\gamma\epsilon )^2b_0^2}
\end{equation}

\begin{figure}[!t]
\includegraphics[clip=true,width=8cm]{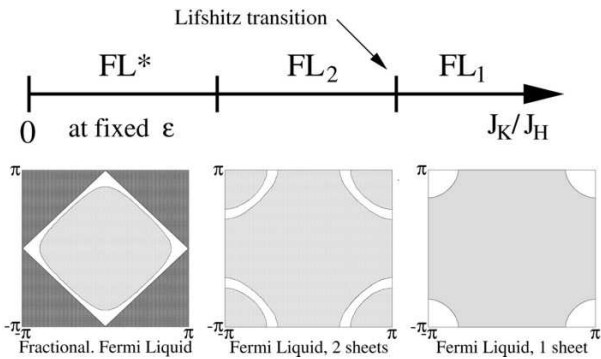}\\
\caption{
\label{sheets}
Fermi surface evolution from FL$^\ast$ to FL,
where shaded areas correspond to occupied states.
Left:    FL$^\ast$, with one spinon (dark) and one conduction electron (light) sheet.
Note that the spinon band is hole-like.
Middle:  FL$_2$, with two sheets, where the outer one represents heavy quasiparticles
with primarily $f$ character.
Right:   FL$_1$, with one heavy-electron sheet.
FL$_2$ and FL$_1$ are separated by a Lifshitz transition where the outer Fermi sheet
disappears at critical value of the ratio $J_H/J_K$.
The conduction band has a filling of $n_c=0.8$.
(The corresponding band structures are also shown in Fig. \ref{singular} below.)
}
\end{figure}

Now consider the FL phase near the transition (small $b_0$). In this case, the two bands
$E_k^\pm$ derive from the $c$ electrons (with weak $f$ character) and the $f$ particles
(with weak $c$ character). For small $b_0$, both bands together therefore intersect the
Fermi energy at least twice so that the Fermi surface consists of two or more sheets
(see Fig. \ref{sheets}).
Upon increasing $b_0$, a topological transition is possible beyond which the
Fermi surface intersects the lower band $E_k^-$ only once and the upper band $E_k^+$ is
empty (Fig.~\ref{singular} below).
This topological splitting of the Fermi surface is a Lifshitz transition.

Let us point out that the existence of this Lifshitz transition in the FL
phase near the Kondo breakdown transition can be seen as a consequence of
the fermionic nature of the spinons in the FL$^\ast$ phase,
which then implies the existence of two fermionic bands near the
FL-FL$^\ast$ transition (also beyond mean-field).
We will comment on other scenarios toward the end of the paper.

\subsection{Numerical results: Phase diagrams}

We have solved the mean-field equations \eqref{largen} and \eqref{epsilon}
numerically for different lattice geometries, namely for electrons
with nearest-neighbor hopping $t$ on 2d square and triangular lattices and
for 3d cubic lattices.\cite{foot5,hermele}
The geometric frustration (here of the triangular lattice) enters our calculation
via the dispersion of both $c$ and $f$ particles.
Note that conventional magnetic order is excluded from the outset due to
the choice of the decoupling scheme.
The lattice degrees of freedom are assumed to be 3d in all cases,
but this does not explicitly enter the mean-field theory \eqref{hmeanfield}.
The calculations are intended for $T=0$, but are performed at a small finite $T$ for
convergence reasons.

\begin{figure}[!t]
\includegraphics[clip=true,width=4.2cm]{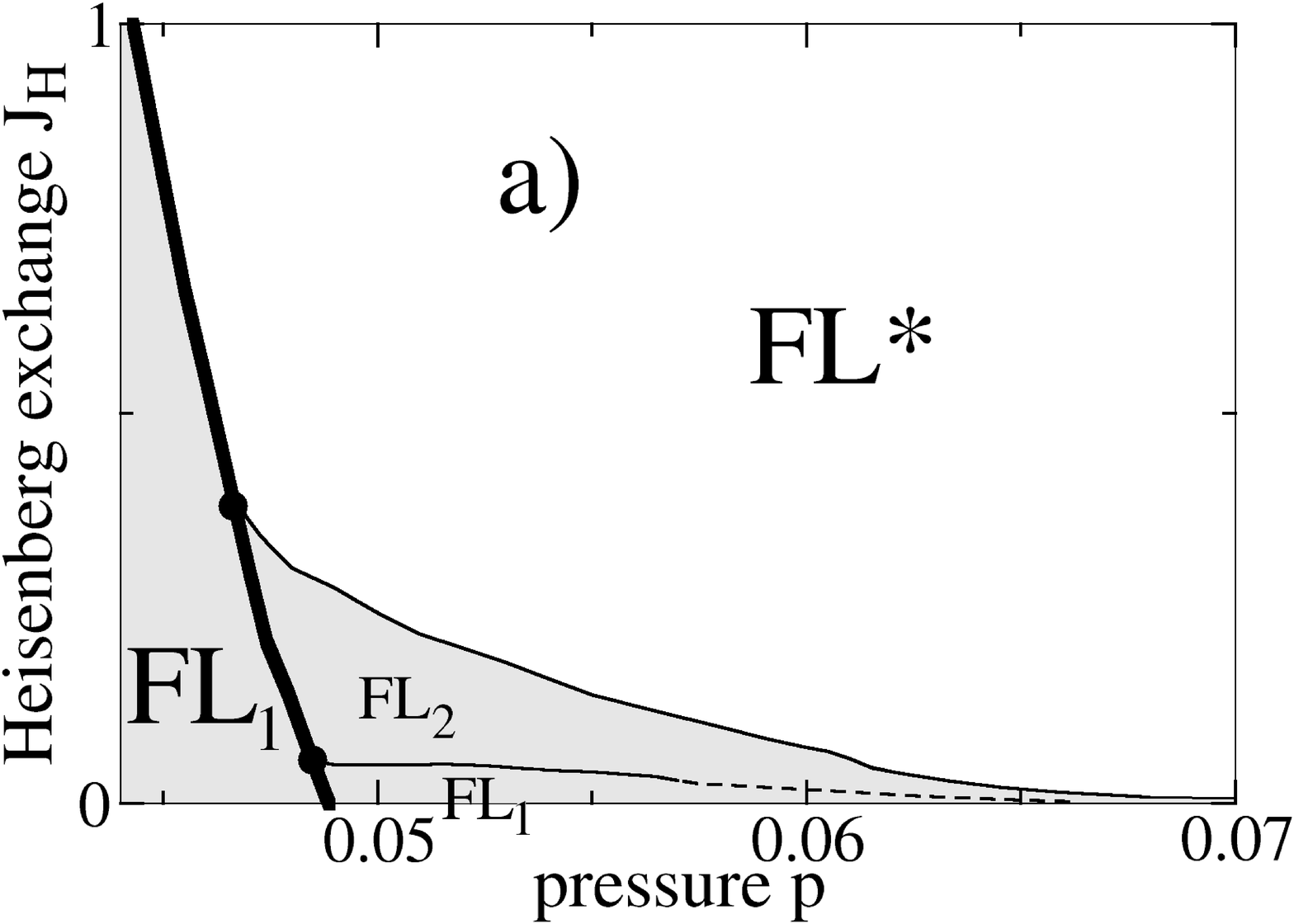}  \includegraphics[clip=true,width=4.2cm]{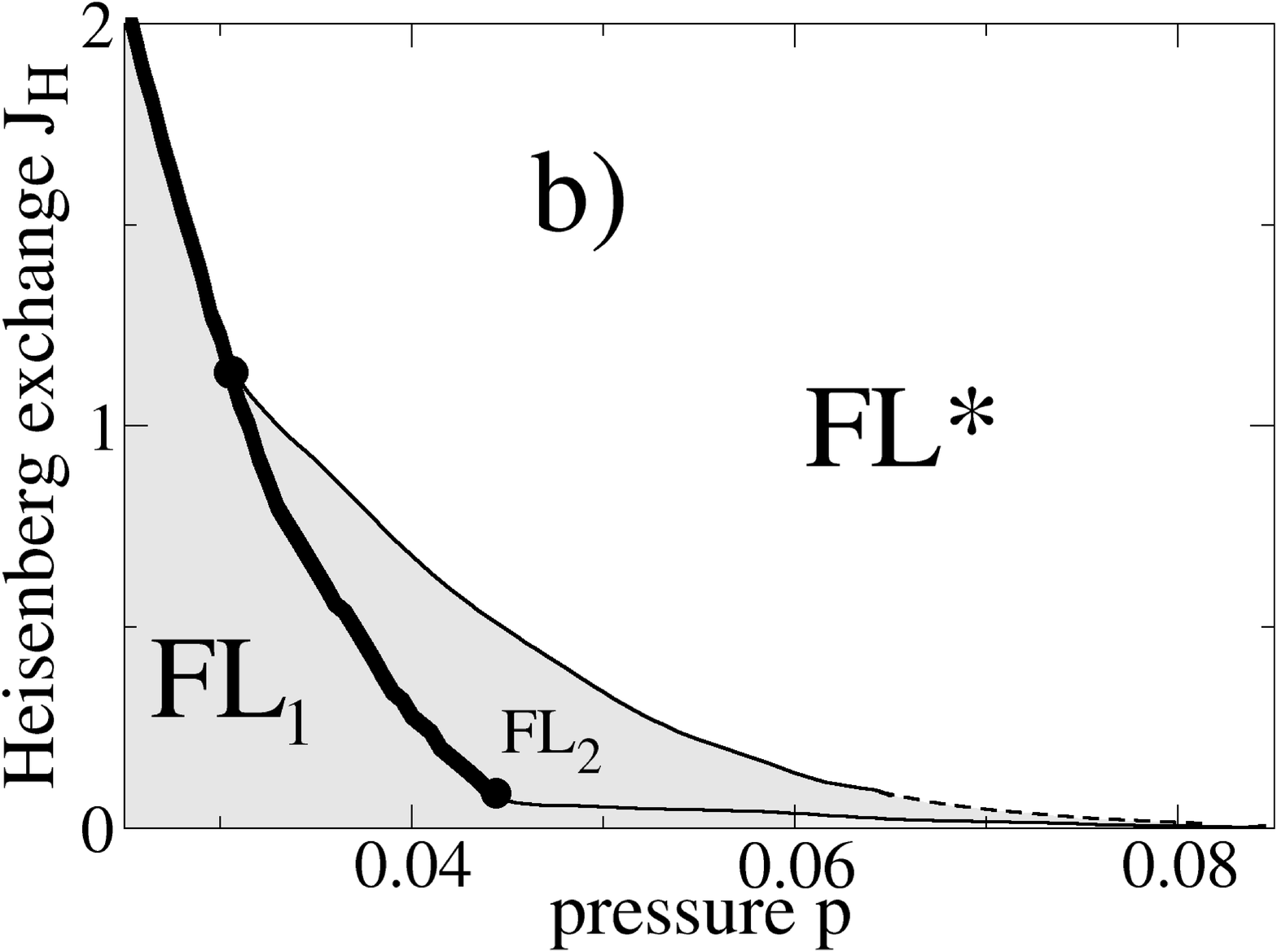}\\
\includegraphics[clip=true,width=4.2cm]{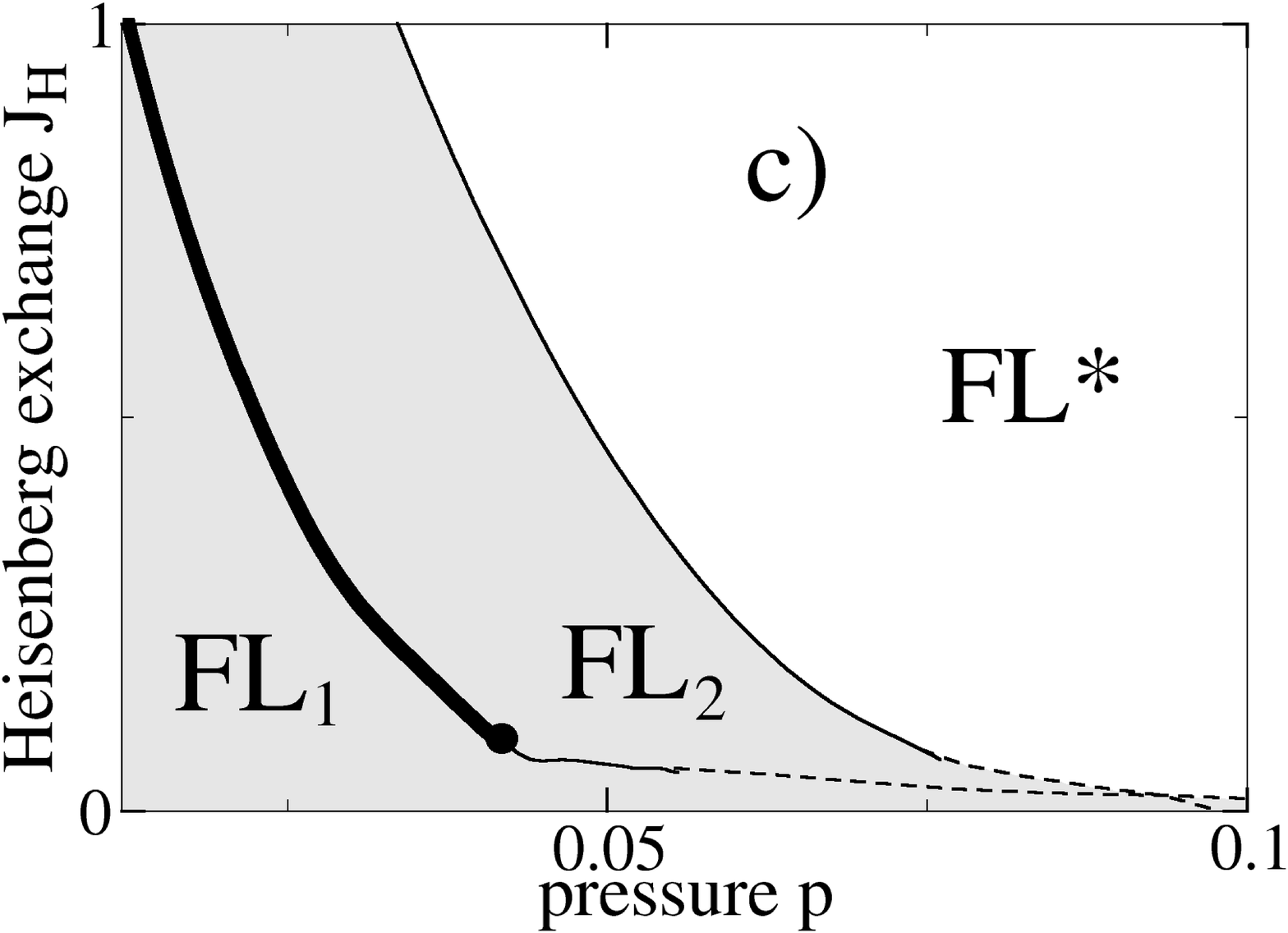}  \includegraphics[clip=true,width=4.2cm]{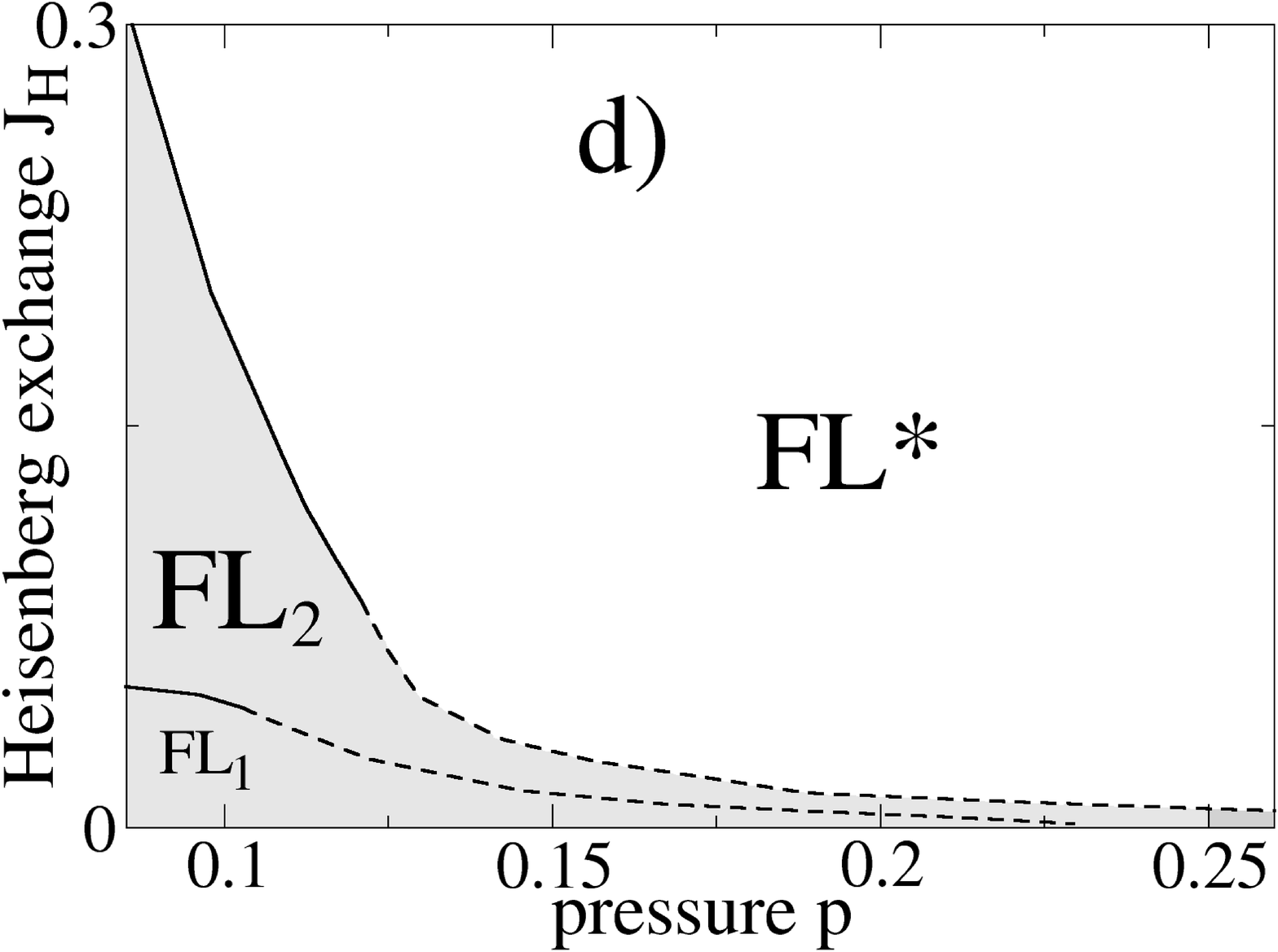}\\
\caption{
\label{2dphase}
Large-$N$ ground state phase diagrams of the Kondo-Heisenberg model \eqref{hmeanfield}
on a 2d square lattice, as function of pressure $p$ and inter-moment exchange $J_H$.
The panels a) to d) are for bulk moduli $B$=0.005, 0.0055, 0.007, and 0.015;
the other parameters are $t$=1, $n_c=0.8$, $J_K=1.5$ and $\gamma=0.05$.
Thin (thick) lines are second (first) order transitions.
The Lifshitz transition line separates FL$_1$ and FL$_2$,
the Kondo breakdown transition separates FL from FL$^\ast$,
for details see text.
(The calculations have been at a low $T=0.005$, the dashed lines are
extrapolations of the phase boundaries obtained from runs at lower $T$.)
}
\end{figure}

We obtain phase diagrams as function of external pressure and inter-moment
exchange, with the bulk modulus $B$ and the conduction band filling $n_c$
as system parameters ($n_f=1$ is fixed in the Kondo limit).
Most qualitative features are independent of the spatial dimensionality and
underlying lattice geometry,
therefore we discuss the 2d square lattice results unless otherwise noted.

Sample phase diagrams for different values of the bulk modulus are shown
in Fig.~\ref{2dphase}.
A sufficiently soft lattice with a correspondingly small bulk modulus
makes a first-order transition at the Lifshitz line preferable. Below a critical bulk
modulus $B^\star$, this first-order lattice transition has no zero-temperature endpoint
(Fig. \ref{2dphase}a). Instead, it continues to $J_H=0$ which is the
Kondo volume-collapse transition considered in Ref.~\onlinecite{schmalian}.
By increasing the bulk modulus above $B^\star$, the first-order transition shows a
quantum critical endpoint which is quite generically located on the Lifshitz line
within numerical accuracy (Fig. \ref{2dphase}b).
Here, for larger values of the inter-moment exchange $J_H$,
the first-order transition coincides with the transition line
to the fractionalized Fermi liquid.
This type of transition, i.e., a direct first-order transition from a heavy Fermi liquid
to a non-Kondo state dominated by inter-moment exchange
(which is FL$^\ast$ here, but will be antiferromagnetic if frustration is reduced)
also appears in other theories,\cite{ogata}
but is usually not observed experimentally.
By further increasing the bulk modulus, the first-order volume transition becomes weaker.
In particular, the coincidence with the transition to the $\text{FL}^*$ phase is shifted to larger
values of the parameter $J_H$  (Fig. \ref{2dphase}c).
Beyond a critical value of the bulk modulus, first-order transitions
do no longer exist, Fig.~\ref{2dphase}d.
This behavior turns out to be well described by a Landau theory,
to be described in Sec. \ref{sec:landau} below.

As our calculations are performed at small finite $T$, care is required if the
characteristic scales of the problem become small.
If $J_H$ or $T_K$ become of order $T$, the transition lines in the phase diagrams start
to deviate from the zero-temperature solution. From numerical results at lower temperatures
we obtained the asymptotic behavior of the transition lines which is indicated by the
dashed lines in Fig. \ref{2dphase}.

\begin{figure}[!t]
\includegraphics[clip=true,width=5cm]{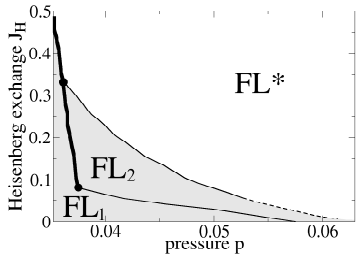}\\
\caption{
\label{3dphase}
As in Fig.~\ref{2dphase}, but for electrons on a 3d cubic lattice,
for $B=0.005$.
}
\end{figure}

The large-$N$ results for 3d electrons, moving on a cubic lattice, are qualitatively similar to
those for $d=2$. An endpoint of the first-order lattice transition line is
found (Fig. \ref{3dphase}) -- whether this endpoint can again be exactly located on the
Lifshitz line without fine tuning is difficult to decide, due to limited numerical resolution.
(This question shall be investigated using the Landau theory of Sec. \ref{sec:landau}
below, with a negative answer.)
Above a critical bulk modulus, the first-order volume collapse transition ceases to exist
(not shown).
Below another critical bulk modulus, the $T=0$ endpoint of the first-order transition
disappears.
As in $d=2$, the first-order line becomes a direct FL$_1$--FL$^\ast$ transition
for large values of the inter-moment exchange $J_H$.

\begin{figure}
\includegraphics[clip=true,width=8.7cm]{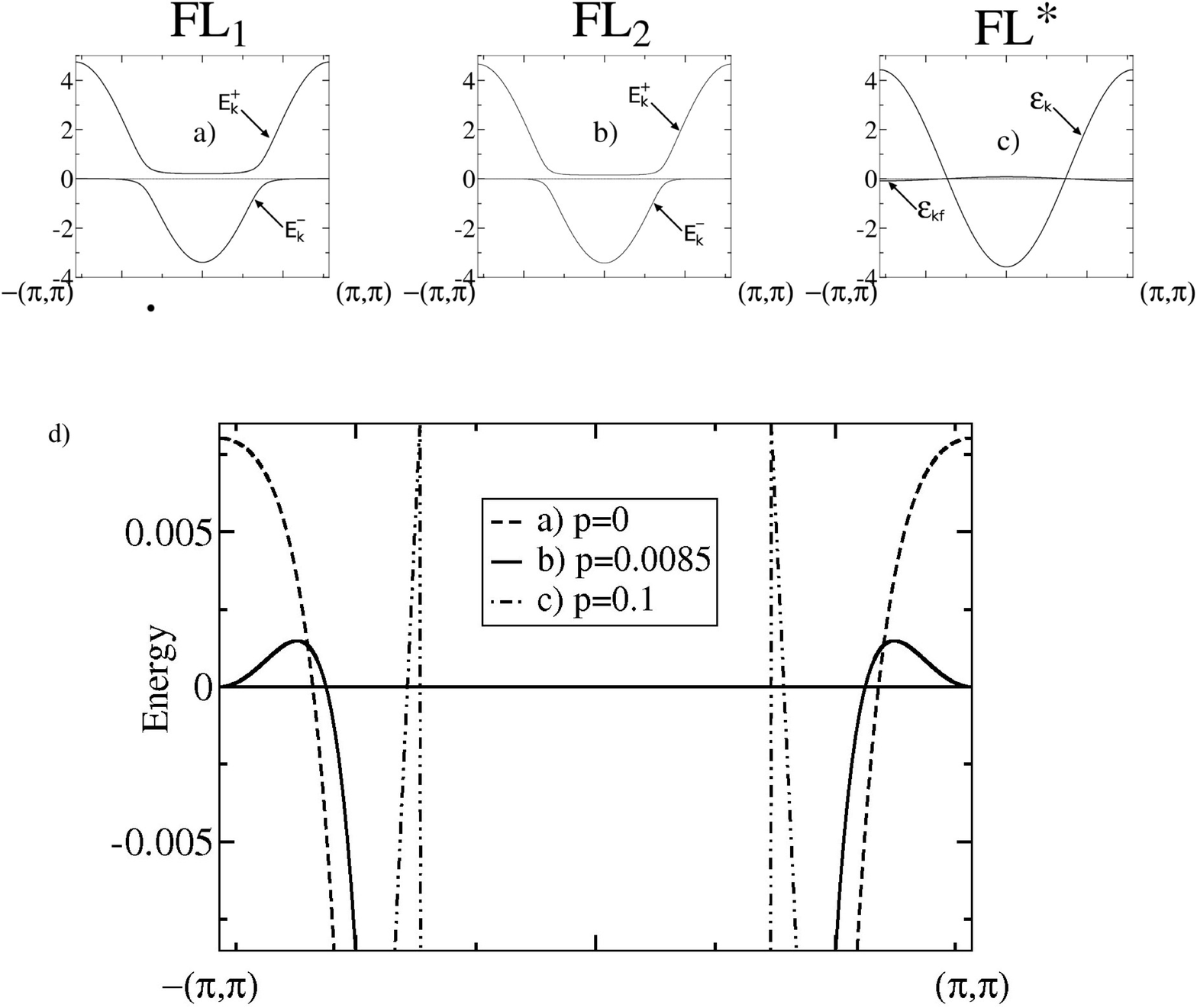}
\caption{
\label{singular}
Quasiparticle dispersions for different parameters along a constant-$J_H$ cut
of the phase diagram in Fig.~\ref{2dphase}, for parameters
$t=1$, $n_c=0.8$, $J_H=0.1$, $J_K=1.5$, $\gamma=0.05$, $B=0.01$ and $T=0$.
Upper panel:
Bands along the $-(\pi,\pi) \rightarrow (\pi,\pi)$ direction for pressures in the
a) FL$_1$, b) FL$_2$ (close to the Lifshitz transition), and c) FL$^\ast$ phases
($p=0$, 0.0085, 0.1, respectively).
Lower panel:
Comparison of the low-energy part of the dispersions.
Case a) has one heavy electron-like band intersecting the Fermi level.
Increasing pressure lowers the energy at the zone boundary and eventually
causes the emergence of a hole-like sheet, case b).
At higher pressures, the two bands evolve into conduction electron and spinon bands,
which each intersect the Fermi energy once - this is the FL$^\ast$ phase, case c).
}
\end{figure}

In our numerical results the mean-field bond amplitude $\chi_0$
is in all aforementioned cases negative, leading to a hole-like spinon dispersion  (see
also Fig. \ref{singular}). At a Lifshitz transition, the lower band $E_k^-$ becomes
suddenly hole-like, promoting a hole-like pocket which splits from the Fermi surface.
Obviously, the emerging hole-like pocket will contribute a power-law density of states
at the Fermi energy, leading to a singular contribution to the thermodynamic potential --
this will be discussed in more detail in the next section.

A comment on the finite-temperature physics is in order:
Both the Lifshitz and Kondo breakdown transitions continue to $T>0$ at mean-field level,
however, it is clear that both are smeared into crossovers upon inclusion of fluctuation
effects.\cite{senthil2}
The first-order lattice transition (for fixed $p$, $J_H$) ends in a
finite-temperature critical endpoint. These endpoints define a line in a generalized
phase diagram which terminates in the $T=0$ endpoint described above.

Let us summarize the physical picture obtained so far.
Upon inclusion of lattice degrees of freedom, the Lifshitz transition
between FL$_1$ and FL$_2$ tends to become first order
(because the system can gain energy by abruptly gapping one of the bands),
while the Kondo breakdown transition remains of second order, except for
soft lattices where both transitions merge into a single first-order transition.
The calculations further suggest an interesting coincidence of two critical
phenomena in the case of 2d electrons,
namely the endpoint of the volume collapse is
located precisely on the Lifshitz transition line.
To understand the latter behavior in more detail, we now devise an
effective theory for the two phenomena.


\section{Landau theory for Lifshitz and lattice volume transitions}
\label{sec:landau}

Two prominent features of the heavy-fermion model of Sec.~\ref{sec:model}
were (i) Lifshitz transitions, where the Fermi surface topology changes with the
appearance or disappearance of a Fermi sheet, and
(ii) lattice transitions of the volume-collapse type.
Note that Lifshitz transitions not only appear in the Kondo-breakdown scenario
of Ref.~\onlinecite{senthil2}, but also in scenarios for metamagnetic behavior
inside the heavy Fermi liquid.\cite{kusmin}
These findings motivate to analyze an effective Landau-type theory for Lifshitz plus
lattice physics --
the mean-field analysis requires less numerical effort than that of the full fermionic
large-$N$ theory of Sec.~\ref{sec:model}.
The discussion of fluctuation effects will be postponed to Sec.~\ref{sec:fluct}.

\subsection{Landau functional}

The vantage point is a conventional Landau theory  of an iso-structural volume-changing
transition of the crystal lattice -- this transition is of liquid--gas type, with
a strain order parameter $\epsilon$, Eq.~\eqref{defeps}.
Further, we assume that the system has a Lifshitz transition, where
the chemical potential crosses a band edge.
Importantly, the lattice parameter $\epsilon$ couples linearly to the tuning
parameter of the Lifshitz transition.
(Microscopically, $\epsilon$ changes e.g. hopping matrix elements, which
in turn moves fermionic bands relative to the chemical potential.)

The band edge crossing the Fermi level gives rise to a {\em non-analytic}
piece in the Landau effective potential at zero temperature.\cite{imada}
The fermion density of states in $d$ dimensions close to the band bottom is:
\begin{equation}
D(\omega) \propto \Theta(\omega) |\omega|^{\frac{d}{2}-1}
\label{dos}
\end{equation}
where $\Theta(x)=0$ (1) for $x<0$ ($x>0$) is the step function.
Then, the non-analytic part of the Landau free energy
arising from the kinetic energy of the fermions is
\begin{equation}
F_{\rm F}(\epsilon ) = - \kappa \Theta (-\epsilon + \epsilon_0 ) \left| \epsilon -\epsilon_0\right|^{\frac{d}{2}+1}
\label{fsing}
\end{equation}
where $\kappa > 0$ is a coupling constant. For $\epsilon<\epsilon_0$ the
band becomes occupied, where $\epsilon_0$ is the location of the Lifshitz transition.

Choosing a reference volume $V_0$ such that $\epsilon_0=0$,
we can add a regular forth-order expansion in the order parameter to obtain
the Landau functional:\cite{landaufoot}
\begin{eqnarray}
F(\epsilon) &=& F_0 - V_0 p^\star \epsilon + \frac{m}{2}\epsilon^2 - \frac{v}{3} \epsilon^3 +
\frac{u}{4}\epsilon^4\nonumber\\
&-&  \kappa \Theta (-\epsilon  )  \left| \epsilon \right|^{\frac{d}{2}+1}
\label{expansion}
\end{eqnarray}
where $p^\star$ is a reference pressure.
Microscopically, the mass $m$ is related to the bulk modulus of the material,
but will also depend on electronic properties (e.g. the ratio $J_H/T_K$ in the Kondo
lattice case).
A cubic term $\epsilon^3$ is symmetry-allowed and generically appears in the
expansion \eqref{expansion}.
(Note that in Ref.~\onlinecite{schmalian} the reference volume was chosen
such that $v=0$ at $T=0$ -- this freedom is absent in the present case.)

The physical situation corresponds to pressure tuning and is described by the
free-enthalpy functional
$G(\epsilon)=F(\epsilon) + pV_0\epsilon$,
which needs to be minimized to obtain the equilibrium value of the strain $\epsilon$.
We shall discuss the arising $T=0$ phase diagram, upon variation of $m$ and $p$,
separately for the cases $d=2$ and $d=3$ in the following subsections.

\begin{figure}
\includegraphics[clip=true,width=3.85cm]{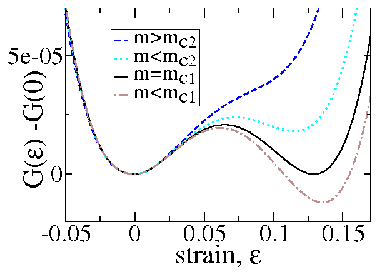}
\includegraphics[clip=true,width=4.15cm]{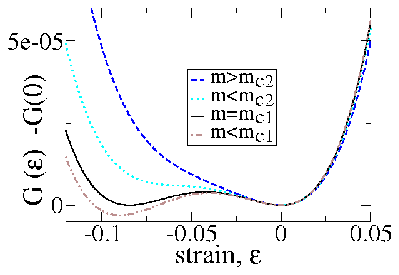}\\
\caption{\label{gibbs}
(Color online)
Samples of the mean-field free energy $F(\epsilon)$ \eqref{expansion},
for various values of the parameter $v$. The left figure
corresponds to $v>0$ and $d=2$, the right figure to $v<0$ and $d=3$. If $m$ drops below
$m_{c2}$, a second minimum evolves at some $\epsilon \neq 0$. If even $m<m_{c1}$, this
minimum becomes the absolute minimum (see also Tables \ref{table1} and \ref{table2}).
}
\end{figure}

Before proceeding, we remind the reader that the thermodynamic singularities at
a Lifshitz transition are in general weak:
The ground-state energy follows $F \propto -\Theta(h) |h|^2$ ($F \propto -\Theta(h) |h|^{5/2}$) for
electrons in $d=2$ ($d=3$) as function of the tuning ``field'' $h$.
In contrast, at a usual second-order transition,
the free energy scales as $F \propto -|h|^{4/3}$ ($F \propto -\Theta(h) |h|^2$)
if $h$ couples linearly (quadratically) to the order
parameter, and we have assumed mean-field exponents
(and neglected fluctuations).
These considerations are relevant, because,
for situations with a linear coupling between order parameter and pressure,
it is known that lattice effects can render the transition first order due to
the compressibility diverging at a finite distance from criticality.\cite{oldelastic}
This effect originates from a diverging second derivative
$-\partial^2 F/\partial h^2$, and does not occur in the Lifshitz case.
In our problem, the compressibility $-\partial^2 G/\partial p^2$ will
only diverge {\em at} the endpoint of the volume collapse transition.

\subsection{Mean-field phase diagram in $d=2$}

By construction, the Lifshitz transition in Eq.~\eqref{expansion} occurs from
$\epsilon>0$ to $\epsilon<0$; a lattice transition may occur elsewhere
on the $\epsilon$ axis.
For $d=2$, the fermionic contribution in Eq.~\eqref{expansion} is quadratic in
$\epsilon$, i.e., renormalizes the mass $m$ for $\epsilon<0$.
In the following, we analyze local minima of the free enthalpy $G(\epsilon)$.
We will obtain critical values $m_{c1,2}$ of the mass $m$:
At $m=m_{c1}$ the Lifshitz transition (as function of $p$) changes from
continuous to discontinuous, and
at $m=m_{c2}$ the lattice transition (as function of $p$) has its critical endpoint.

A continuous Lifshitz transition is not possible if $G(\epsilon)$ is not minimal at
$\epsilon=0$ at any pressure. This is certainly the case if $m-2\kappa < 0$. If
$m-2\kappa \geq  0$, the position of minima depends crucially on sign and magnitude of
the cubic coefficient $v$, as depicted in Fig. \ref{gibbs}. We distinguish these cases in
successive order.

\begin{figure}[!b]
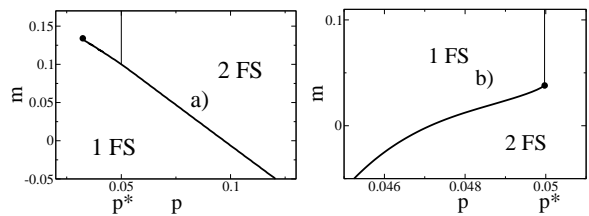

\includegraphics[clip=true,width=3.9cm]{landau.eps}
\includegraphics[clip=true,width=3.75cm]{landau2.eps}\\
\caption{\label{meanfield1}
Two representative mean-field phase diagrams in two dimensions, derived
from the Landau theory.
Thick/thin lines represent first/second order transitions.
If the parameters $m_{c1}$ and $m_{c2}$ coincide, the quantum critical endpoint of the lattice
transition is located on the Lifshitz transition line (Fig. \ref{meanfield1}b).
In the other case, $m_{c1}\neq m_{c2}$, the first-order lattice transition line turns away from the
Lifshitz line and ends at a critical endpoint (Fig. \ref{meanfield1}a).
}
\end{figure}

(i) $v>0$.
Obviously $G(\epsilon>0)$ can acquire a
second minimum if $v$ becomes large such that
\begin{equation}
\frac{dG(\epsilon)}{d\epsilon} = p-p^\star + m \epsilon -v\epsilon^2 +u \epsilon^3 \stackrel{\rm !}{=} 0
\label{cubic}
\end{equation}
is possible if $\epsilon>0$. The existence of two minima of $G(\epsilon)$ is a necessary
condition for a discontinuous transition to occur. The discriminant of the cubic equation
(\ref{cubic}) excludes a second minimum if $v^2 \leq  \frac{9}{2}u m$. Obviously the
first-order transition line terminates then at the critical endpoint $m_{c2}=2\kappa$ as
long as $0 < v \leq  3\sqrt{u\kappa}$.
In the opposite case, $v > 3\sqrt{u\kappa}$, a second minimum of $G(\epsilon>0)$ develops for
pressures that lead to three real solutions of Eq. (\ref{cubic}). If $v >
\frac{9}{2}\sqrt{u\kappa}$, in this second minimum $G(\epsilon)$ becomes smaller than
$F_0$ at $p=p^\star$ (Fig. \ref{gibbs}, left).
The Lifshitz transition changes from first to second order
at $m_{c1}=\frac{2}{9} \frac{v^2}{u}$.
The first-order lattice transition line finally terminates at the critical endpoint
$m_{c2}=\frac{1}{3}\frac{v^2}{u}$ where the second minimum vanishes (Fig.
\ref{meanfield1}).
Note that $m_{c1}=m_{c2}$ if $v \leq v_{c1} = \sqrt{6u\kappa}$, which leads to a critical endpoint
located {\em at} the Lifshitz line (Fig.~\ref{meanfield1}b).

\begin{table}
\begin{ruledtabular}
\begin{tabular}{c| c c c}
$\boldsymbol{d} \boldsymbol{=} \boldsymbol{2}$ & $0 \leq  \left| v \right| \leq v_{c1}$ & $v_{c1} <  \left| v \right| \leq v_{c2}$ &
 $v_{c2}< \left| v \right|$  \\
\hline
$m_{c1}$ & $\frac{2}{9}\frac{v^2}{u}\Theta(-v) +2\kappa$ & $\frac{2}{9}\frac{v^2}{u}\Theta(-v) + 2\kappa$ &  $\frac{2}{9}\frac{v^2}{u}+ 2\kappa\Theta(-v)$\\
$m_{c2}$ & $\frac{1}{3} \frac{v^2}{u}\Theta(-v)+ 2\kappa$ & $\frac{1}{3}\frac{v^2}{u}+ 2\kappa\Theta(-v)$ & $\frac{1}{3}\frac{v^2}{u}+ 2\kappa\Theta(-v)$ \\
\end{tabular}
\end{ruledtabular}
\caption{\label{table1}
Location of critical points from Landau theory in $d=2$,
in dependence of the cubic coefficient $v$.
The critical values are $v_{c1}= \sqrt{6u\kappa}$ and $v_{c2}= 3\sqrt{u \kappa}$.
If $m_{c1}<m_{c2}$, a critical endpoint $m_{c2}$ away from the Lifshitz transition emerges.
This is not possible if $0 \leq v \leq v_{c1}$  since always $m_{c1}=m_{c2}=2\kappa$ then.
See text for details.
}
\end{table}

(ii) $v<0$.
Our analysis of positive coefficients $v$
does not need to be modified if we repeat all arguments for an order parameter $\epsilon<0$.
In this case the Lifshitz transition always becomes continuous at
$m_{c1}=\frac{2}{9}\frac{v^2}{u}+2\kappa$,
and the first-order lattice transition line terminates at the critical
endpoint $m_{c2}=\frac{1}{3}\frac{v^2}{u}+ 2\kappa$.
The phase diagram corresponds then always to Fig. \ref{meanfield1}a.

(iii) $v=0$.
In this trivial case the critical
endpoint $m_{c1}=m_{c2}=2\kappa$ is situated on the Lifshitz line.

We summarize all different cases in two spatial dimensions in Table \ref{table1}.
Qualitatively, all these cases divide into two different types of phase diagrams as
depicted in Fig.~\ref{meanfield1}.

If the critical endpoint is on the Lifshitz line, then the
effective potential at this special point is $\propto |\epsilon|^2$
($\propto |\epsilon|^3$) for positive (negative) $\epsilon$.
This implies that $\epsilon \propto (p-p^\ast)^{1/2}$ ($\epsilon \propto p^\ast-p$)
for $p>p^\ast$ ($p<p^\ast$) in the vicinity of the endpoint, i.e., for $m=m_{c1,2}$.
The first-order line near this point follows $|m-m_{c1,2}| \propto |p-p^\ast|^{1/2}$,
and along this line the jump in $\epsilon$ is $\Delta\epsilon \propto |m-m_{c1,2}|$.
This shows that the critical exponents $\beta$ and $\delta$ do not take
their mean-field values, due to the non-analytic behavior of the fermionic part of the
Landau potential.

\subsection{Mean-field phase diagram in $d=3$}

In three dimensions, the singular contribution to the free energy enters with a
fractional power which is not contained in the series expansion of the regular part.
In addition to the critical parameters $m_{c1}$ and $m_{c2}$, it is necessary to introduce
further critical parameters $m_{c1}^\prime(u,\kappa,v)$ and $m_{c2}^\prime(u,\kappa,v)$
which mark the onset of local minima of $G(\epsilon)$ at $\epsilon<0$
(Fig. \ref{gibbs}b).
If pressure is used as control parameter, the relation of $m$ to the
critical values $m_{c1}$, $m_{c2}$, $m_{c1}^\prime$ and $m_{c2}^\prime$ determines the
type of phase transition. If $m < 0$, generically a discontinuous Lifshitz transition
occurs at a critical pressure $p>p^\star$. Furthermore, if $v \leq 0$ we obtain
$m_{c1}=m_{c2}=0$. By varying the parameter $\kappa$ and thereby tuning $m_{c1}^\prime$
and $m_{c2}^\prime$, all entries of Table \ref{table2} can be realized.

\begin{table}
\begin{ruledtabular}
\begin{tabular}{c| c c c}
 $\boldsymbol{d} \boldsymbol{=} \boldsymbol{3}$ & $0 \leq  m < m_{c1}$ & $m_{c1}\leq  m <m_{c2}$ &  $ m_{c2}\leq  m$ \\
\hline
$0 \leq  m < m_{c1}^\prime$ & $1^{\text{st}}$ \& LS& $1^{\text{st}}$ \& LS \footnotemark[1] &  $1^{\text{st}}$ \& LS\\
$ m_{c1}^\prime \leq  m <m_{c2}^\prime $ & $1^{\text{st}}$ \& LS\footnotemark[1] & $1^{\text{st}}$\footnotemark[2] &  $1^{\text{st}}$ \\
$ m_{c2}^\prime\leq m $  & $1^{\text{st}}$ \& LS & $1^{\text{st}}$ &  $2^{\text{nd}}$  \\
\end{tabular}
\end{ruledtabular}
\footnotetext[1]{an additional $1^{\text{st}}$ order transition can occur}
\footnotetext[2]{two $1^{\text{st}}$ order transitions occur}
\caption{\label{table2}
Type of phase transitions from Landau theory in $d=3$,
for $m \geq 0$ and $v > 0$. We denote a
first-order transition which is also a Lifshitz transition by the entry
$1^{\text{st}}$ \& LS. If $m > 0$ and $v \leq 0$, we know from the 2d case
that $m_{c1}=m_{c2}=0$ and only the third column can be realized. The latter case leads
again to the phase diagram of Fig. \ref{meanfield1}b. In the opposite case $v > 0$ we
always obtain $m_{c2}>m_{c1}>0$. From the above table we conclude then that in addition
to Fig. \ref{meanfield1}b, also the phase diagrams of Fig. \ref{meanfield2} can be
realized.
}
\end{table}

\begin{figure}[!t]
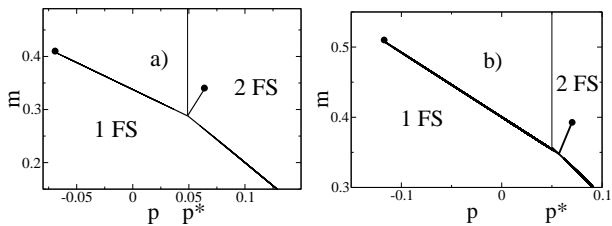

\includegraphics[clip=true,width=4.0cm]{landau4.eps}
\includegraphics[clip=true,width=4.0cm]{landau3.eps}\\
\caption{\label{meanfield2}
Additional cases for mean-field phase diagrams that occur in $d=3$.
Both situations the first-order lattice transition line bifurcates
at a point which a) does [b) does not] coincide with the point where
the Lifshitz transition changes from first to second order.
}
\end{figure}

From Table \ref{table2} we can read off the phase diagrams possible in $d=3$.
Since always $m_{c1}^\prime<m_{c2}^\prime$, the zero-temperature endpoint
of the lattice transition is never located on the Lifshitz line (except with
further fine-tuning).
The behavior along the Lifshitz line can be again of the type shown in
Fig.~\ref{meanfield1}a, or there may be additional bifurcations of the
first-order lattice transition (Fig.~\ref{meanfield2}).

\subsection{Conclusions from Landau theory}

It has been shown that, for a range of mass parameters $m$, the Lifshitz
transition is rendered discontinuous due to a coupling to the lattice, i.e.,
a first-order volume collapse coincides with the Lifshitz transition.
The volume collapse displays a quantum critical endpoint in the $m$--$p$ plane.
As a specific feature of the $d=2$ theory, the volume-collapse endpoint is located
{\em on} the Lifshitz transition line in a large parameter regime.
(In contrast, in $d=3$ the Lifshitz and volume-collapse transitions separate before the
volume-collapse endpoint is reached.)


\section{Beyond mean-field theory}
\label{sec:fluct}

So far, our analysis was restricted to mean-field theory derived either from a
large-$N$ or a Landau approach.
In this section, we briefly comment on fluctuation effects near the continuous
quantum phase transitions of our model system.
(It is reasonable to assume that discontinuous mean-field transitions remain of
first order up to a critical temperature where a finite-temperature endpoint
of liquid--gas type occurs.)

The second-order Kondo breakdown transition has been investigated in
Refs.~\onlinecite{senthil2} and \onlinecite{paul} -- this can lead to interesting
non-Fermi liquid behavior, but we shall not repeat the arguments here.
The Lifshitz transition is in the universality class of the dilute Fermi gas,
with finite-temperature properties discussed e.g. in Ref.~\onlinecite{ssbook}.

It remains to consider the endpoint of the volume-collapse transition.
This has a remarkable feature:
The critical endpoint of an iso-structural volume-collapse transition has
{\em no} critical fluctuations at finite wavevectors.
To see this, first note that the strain field $\epsilon$ is related
to longitudinal phonons by $\epsilon \sim \nabla_x \cdot \phi$
where $\phi$ is the displacement field.
In continuum elasticity theory, it is well known that, even for a
vanishing bulk modulus, a finite shear modulus renders all phonon velocities
finite.\cite{LL}
A detailed analysis by Cowley\cite{cowley} shows that the same is true in
crystals: All homogeneous strain fluctuations remain non-critical at the
iso-structural volume-collapse transition.
(Note that the vanishing of a phonon velocity would imply a structural
transition with symmetry breaking.)
In other words, at the endpoint under consideration,
only the zero-momentum mode becomes critical, and Landau theory is exact
for the lattice degrees of freedom.\cite{cowley}
(Strictly speaking, phonon modes with wavelengths of order sample size
become critical.)

For the interesting case of coinciding volume-collapse endpoint and Lifshitz
transition, we can briefly discuss the thermodynamics.
To this end, temperature corrections to the mean-field parameters $p^\ast$
and $m$ of Eq.~\eqref{expansion} have to be taken into account.
Due to the underlying Fermi-liquid physics (of the additional bands crossing
the Fermi level), these are quadratic in temperature,
e.g., $p^\ast = p^\ast(T=0) + \zeta_1 T^2$.
Furthermore, the fermionic contribution to Eq.~\eqref{expansion}
has to be replaced by the free energy of the Fermi gas.
Near the Lifshitz transition this can be written as\cite{ssbook}
\begin{equation}
F_{\text{F}}(\epsilon,T) = - \kappa \, T^{\frac{d}{2}+1} \, \Phi(\epsilon/T)
\end{equation}
which replaces Eq.~\eqref{fsing}, and
$\Phi(x)$ is a scaling function.
Minimizing the free enthalpy $G(\epsilon)$ in $d=2$ now shows that
$\epsilon \propto T$ upon raising temperature at the endpoint, i.e.,
the behavior is dominated by the fermionic part of the free energy.
Consequently, the compressibility diverges as $\kappa \propto T^{-1}$,
and the specific heat follows $C_p \propto T$.


\section{Discussion}
\label{disc}

Our study of coupled Kondo-lattice and volume-fluctuation physics
was motivated by antiferromagnetic quantum critical points in heavy-fermion
metals where the proposal of a breakdown of the Kondo effect at criticality
has spurred intense interest.
Within the scenario of Ref.~\onlinecite{senthil2},
where a transition between FL and FL$^\ast$ occurs via the vanishing of the
quasiparticle weight on one of the two sheets of the FL Fermi surface,
we have investigated whether first-order volume transitions may spoil
quantum criticality.
These calculations merge former investigations on Kondo breakdown and
Kondo volume collapse transitions.

Our results show:
(i) The Kondo breakdown transition remains intact as a second-order transition
(except for very soft lattices),
but the Lifshitz transition (i.e. a topological splitting of the Fermi surface)
which inevitably appears near the FL--FL$^\ast$ transition in the scenario of
Ref.~\onlinecite{senthil2}, tends to become first-order, i.e.,
occurs concommitantly with a Kondo volume collapse.
(ii) Applied to the heavy-fermion metals \CeAu\ and \YbRhSi,
this suggests to experimentally search for (possibly weak) first-order
transitions at low temperature near the unconventional candidate critical
points.
If those are not found, then alternative scenarios of Kondo breakdown
should be investigated in more detail\cite{coleman,si} --
those may not necessarily imply the simultaneous existence of two Fermi
sheets near the Kondo breakdown transition.
Particularly appealing seems the idea that, at the transition point,
quasiparticles on both the large Fermi surface of FL and the small Fermi surface
of FL$^\ast$ (or of the ordered state) become singular, i.e., that two
hot Fermi surface sheets coexist, with a ``super-large'' Fermi surface of
volume $1+2n_c$.\cite{senthilpriv}
(As a result, the Lifshitz transition is removed from the agenda.)
However, a mean-field theory for this scenario is not known.
(iii) As an aside, we found that the zero-temperature endpoint coincides
with the second-order Lifshitz transition for a large parameter regime in
the case of 2d electrons.
This coincidence of two critical phenomena, i.e. ``non-Landau behavior'',
is rooted in the non-analytic free energy near the topological Lifshitz
transition.


\begin{acknowledgments}
We are indebted to M. Garst and A. Rosch for various illuminating discussions,
especially on the role of fluctuations at structural phase transitions.
Furthermore, we acknowledge helpful conversations with J. Schmalian and
T. Senthil.
This research was supported by the DFG through the SFB 608 (K\"oln)
and the Research Unit FG 960 ``Quantum Phase Transitions''.
\end{acknowledgments}



\begin{thebibliography}{99}
\bibitem{foot2}
We consider the particle-hole symmetric case $\epsilon_f=-\frac{U}{2}$ here,
and neglect possibly arising potential-scattering terms in the Kondo model.
\bibitem{foot3}
Ref.~\protect\onlinecite{paul}
discusses an additional modulated mean-field solution which may
occur when the band masses of electrons and spinons have opposite sign.
In this phase, $b_0$ is not spatially uniform which implies a
charge density modulation in the heavy Fermi liquid.
To our knowledge, such a phase is not observed experimentally.
\bibitem{foot5}
The U(1) FL$^\ast$ is established as a stable phase in $d=3$ dimensions,\cite{senthil2}
its stability in $d=2$ is a matter of current discussion.\cite{hermele}
However, even if FL$^\ast$ is unstable at lowest temperatures, the transition between
FL and FL$^\ast$ can be relevant for the critical behavior over an intermediate
range of energies or temperatures.\cite{senthil2,senthil3}
(On the level of the mean-field calculations, the distinction between 2d and 3d is
only in the dispersion.)
\bibitem{landaufoot}
In principle, the parameters of the Landau expansion, Eq.~\eqref{expansion},
can be related to those of the microscopic Kondo lattice model
of Sec.~\ref{sec:model}, by expanding the mean-field free energy
around the numerical mean-field solution at the Lifshitz transition.
\end{thebibliography}
\end{document}